\documentclass[Crown,sageh,times]{sagej}

\usepackage{moreverb,url}

\usepackage[colorlinks,allcolors=red]{hyperref}

\newcommand\BibTeX{{\rmfamily B\kern-.05em \textsc{i\kern-.025em b}\kern-.08em
T\kern-.1667em\lower.7ex\hbox{E}\kern-.125emX}}

\def\volumeyear{2020}

\newcommand{\X}{\mathbf{X}}

\begin{document}

\runninghead{Vacant Lot Greening in Philadelphia}

\title{The Effects of Vacant Lot Greening and the Impact of Land Use and Business Presence on Crime}

\author{Jesse Cui\affilnum{1} and Shane T. Jensen\affilnum{2} and John Macdonald\affilnum{3}}

\affiliation{\affilnum{1}Wharton Social Impact Initiative, The Wharton School, University of Pennsylvania\\
\affilnum{2}Department of Statistics, The Wharton School, University of Pennsylvania\\
\affilnum{3}Department of Criminology and Sociology, University of Pennsylvania}

\corrauth{Shane T. Jensen, 
Department of Statistics,
The Wharton School,
University of Pennsylvania,
463 Huntsman Hall
3730 Walnut Street,
Philadelphia, PA, USA 19102}

\email{stjensen@wharton.upenn.edu}

\begin{abstract}
We examine the effect of the Philadelphia LandCare (PLC) vacant lot greening initiative on crime and the extent to which surrounding land uses and business types moderate this intervention.  We rely on a propensity score matching analysis to account for substantial differences in demographic, economic, land use, and business characteristics between greened and ungreened vacant lots.  We estimate larger and more significant crime reductions around vacant lots that are greened in our matched pairs analysis compared to unmatched analyses.  The effects of vacant lot greening on crime are larger in areas with high residential and low commercial land use and are moderated by the presence of different types of nearby businesses.  
\end{abstract}


\maketitle

\section{Introduction}

The recent availability of urban data provides the opportunity to investigate features of neighborhood environments at a higher resolution than ever before and quantitatively test how urban design principles are related to human activities.  One important area of research is examining how the characteristics of the built environment can impact the public safety of local neighborhoods.  

We focus on a particular type of built environment intervention: the ongoing Philadelphia LandCare (PLC) vacant lot greening program carried out by the Pennsylvania Horticultural Society.  In this ongoing intervention, thousands of vacant lots in the city of Philadelphia have been cleaned up and turned into small public spaces in an effort to improve the surrounding area.   Our objective is to estimated the effects of this vacant lot greening program on crime in recent years.  Our fine grained analysis compares lots that are remediated to those that remain vacant and unkempt but are located in similar surrounding land use contexts. We also assess the extent to which the effects of vacant lot greening on crime are moderated by surrounding land use and the presence of different types of businesses. This work extends earlier research by \cite{macdonald2021reducing} that compared changes in crime among vacant lots that were greened and those that remained vacant within the same census tract.  We undertake a more direct matching of greened and ungreened lots based on their surrounding economic, demographic, and land use characteristics.  Our approach assures that greened and ungreened vacant lots are directly comparable on nearby neighborhood context.  We also extend the work of \cite{macdonald2021reducing} by examining whether the presence of specific businesses types impacts the effects of vacant lot greening on crime.  This investigation provides a more detailed understanding of the specific role that nearby businesses play as proprietors of public space and their impact on crime. 

\subsection{Theories relating the Built Environment and Safety}\label{theories}

There are many theories in urban planning and criminology that hypothesize associations between aspects of the built environment, human activity and safety.  A seminal idea by Jane Jacobs in her 1961 book, The Death and Life of Great American Cities, is the concept that the built environment can encourage social activities through commercial and residential uses that generate more ``eyes upon on the street" \citep{Jac61}.  The Defensible Space theory of \citep{New72} also argues that the architecture of buildings and the surrounding built environment can reduce crime by stimulating more natural surveillance of public activities and the perception of territorial control over public spaces. 

These ideas were expanded into the theories of crime prevention through environmental design \citep{Jef71,CozSavHil05}, situational crime prevention \citep{Cla95}, human territorial functioning \citep{Tay88}, and collective efficacy \citep{Sam12}, all of which hypothesize that the built environment can foster signals that a place is more or less attractive for committing crimes. Crime prevention through environmental design and situational crime prevention argue that the built environment sends signals to offenders about the level of guardianship and surveillance in a location \citep{WilCul18}.  For example, an overgrown and unkempt vacant lot may suggest there are few proprietors of this space which would make it easier to engage in criminal activity and avoid detection.  Collective efficacy and human territorial function both suggest that abandoned and overgrown vacant lots, along with other signs of physical disorder, disrupt the social fabric of a block through the perception that no one is maintaining control of the space as well as undermining the collective actions of residents to engage in informal social controls that help mitigate criminal behavior.  
These theoretical perspectives lead to the expectation that crime should be reduced around vacant lots if they are greened and maintained. There is also empirical support of this expectation from previous studies on this PLC initiative \citep{branas-1, MoyMacRid19, Macbrasto19}, which we discuss in the next section.  

At the same time, the surrounding land use of places is also a neighborhood feature that may moderate any built environment intervention, including the cleaning and greening of vacant lots. For example, commercial land use and business presence that increases the number of both perpetrators and potential victims in an area may diminish the crime reducing effects of greening vacant lots.  Known crime generators like retail businesses \citep{BerBlo09} that do not act as proprietors of their nearby spaces (especially if these spaces are vacant) may undermine the benefits of vacant lot remediation.  On the other hand, certain businesses and residents may be more likely to engage in place management \citep{WilCul18} of nearby vacant lots after they have been cleaned and greened. Thus, businesses and residents may act as proprietors of vacant lots once they have been remediated.

In Section~\ref{land-use-business-results}, we will investigate these hypotheses by evaluating the impact of surrounding land use zoning and business presence on the effects of the Philadelphia LandCare vacant lot greening initiative on crime.  

\subsection{Empirical Evaluations of the Built Environment and Vacant Lot Greening}

Historically, theories about the impact of the built environment have been difficult to test because our ability to experiment with changes in urban environments is severely limited by the challenges of purposively altering the city environment and imposing treatments on a human population.  However, in the past ten years a growing body of research has employed experimental and quasi-experimental designs to examine what happens to crime after there have been changes to public space, housing, transit, and zoning \citep{Mac15, Macbrasto19}.  More specific to greening interventions, previous empirical evidence has found that the cleaning and greening of vacant lots is associated with reductions in violent crimes in Youngstown, OH and Flint, MI, as well as drug crimes in New Orleans, LA \citep{KonHohHan16,KonMorJac18,HeiKruVag18}.   

\cite{branas-1} examined the early years of the PLC vacant lot greening program that is also the focus of our study.  They used a difference-in-differences analysis of the impact of vacant lot greening in Philadelphia and found reductions in gun assaults and vandalism.  A smaller randomized-control trial of the PLC greening program showed reductions in serious crime, fear of crime, and shootings \citep{BraSouKon18,MoyMacRid19}.   More recently, \cite{macdonald2021reducing} examined whether the effects on crime were moderated by nearby train stations, alcohol outlets, the count of active businesses, or nearby land that is zoned for commercial or mixed uses.  They find evidence that PLC greening leads to smaller reductions in crime when remediated lots are nearby trains or alcohol outlets.  

In our contribution outlined below, we more directly assess the role that surrounding land use and specific business types have on conditioning the effects of vacant lot greening on crime.  

\subsection{Our Contribution}

The Philadelphia LandCare vacant lot greening program provides a unique quasi-experimental setting where this built environment intervention can be evaluated by comparing the change in crime incidence around greened versus ungreened vacant lots.   However, when performing these comparisons, we must be extra careful to take into account the surrounding neighborhood context of these vacant lots.  Systematic differences between greened and ungreened vacant lots in terms of their surrounding characteristics can confound these comparisons since observed differences in crime cannot be attributed to the greening intervention.  We will see in Section~\ref{data-comparisons} that the areas surrounding greened vacant lots are substantially different from the areas surrounding unkempt vacant lots in terms of their demographic, economic, and land use characteristics.

We address these systematic differences in surrounding context between greened and ungreened lots by matching lots based on their surrounding neighborhood context.  We use propensity score matching \citep{RosRub83} to create matched pairs of vacant lots where each pair consists of one greened lot and one ungreened lot that have highly similar demographic, economic, and land use characteristics.   These matched pairs allow us to make more balanced comparisons of crime rates between greened and ungreened lots in order to provide a comprehensive evaluation of the effects of this vacant lot greening intervention in Philadelphia in recent years. 

Beyond this overall evaluation, we are particularly interested in whether the effect of vacant lot greening on crime is conditioned by nearby land use and business presence.  Rather than assess moderation of PLC program based on the overall number of active businesses \citep{macdonald2021reducing}, we examine whether there are specific types of businesses that serve as crime generators or place managers that either undermine or enhance the benefits of vacant lot greening on crime.  

In addition to demographic and economic data from the U.S. Census Bureau, we use zoning data from the City of Philadelphia to create detailed measures of land use around each greened and ungreened vacant lot.  We also incorporate detailed business location data from 
\cite{humphrey2020urban} into our analysis.  All of these data sources will be involved in the creation of our matching procedure in order to ensure our matched pairs of greened compared to ungreened lots are similar on many different aspects of their surrounding context.  

These matched pairs also facilitate our investigation of the influence of nearby land use and business types on the effects of vacant lot greening on crime.   Since each matched pair will contain two lots with highly similar land use and business presence characteristics, we can subset our pairs in order to explore whether the effect of vacant lot greening on crime changes between pairs that differ substantially on aspects of their surrounding land use or types of businesses.  

In summary, we harness sophisticated matching methodology and available high resolution data in Philadelphia to evaluate the effects of the PLC vacant lot greening program on crime and whether nearby land use or business presence influences the impact of this greening intervention.   We estimate an extra 1-2\% reduction in total crimes per year around greened vacant lots compared to their matched ungreened vacant lots.  We also find a significant role for surrounding land use: the effects of vacant lot greening seem most beneficial in areas with high residential and low commercial zoning, which is consistent with several theories outlined in Section~\ref{theories}.  We also find significant impacts for the presence of several business types that suggest some types of businesses act as proprietors of these public spaces while others may be crime generators.

In Section~\ref{data-comparisons}, we describe our data for Philadelphia and compare crime between the unmatched sets of all greened and ungreened vacant lots, while also observing systematic differences in their surrounding neighborhood context.  We address these systematic differences with a careful matching of greened vs. ungreened vacant lots and use our matched pairs to evaluate the effect of vacant lot greening on crime in Section~\ref{matching}.   In Section~\ref{land-use-business-results}, we use different subsets of these matched pairs to investigate the impact of different types of land use and business presence on the effect of vacant lot greening on crime.  We conclude with a brief summary and discussion in Section~\ref{discussion}.   The code repository for data processing and analysis can be viewed at https://github.com/jessecui/WSII-Urban-Analytics-Business-Vibrancy.

\section{Urban Data and DID Comparison of Greened vs. Ungreened Vacant Lots}\label{data-comparisons}

Our analyses will be based on publicly available data on crime, economic, and demographic neighborhood characteristics, land use zoning and business locations.  
We have the location and type of every reported crime over the past decade from the Philadelphia Police Department.  We categorized these $\approx 1.5$ million reported crimes into two major types: \textbf{serious crimes} and \textbf{other crimes}.  Serious crimes include homicides, rapes, robberies, aggravated assault, and other assaults.  Other crimes are a composite of both property and nuisance crimes.   Details of our crime categorization is provided in our supplementary materials.  

We also have detailed data on neighborhood-level income, poverty, population, and race from the U.S. Census Bureau.   Land use data from the City of Philadelphia provides the zoning designation for $\approx 560,000$ lots in the city, which we categorize into eight types of land use: Residential, Commercial, Industrial, Civic, Transportation, Cultural, Water, and Vacant. Our research group also compiled a comprehensive database of Philadelphia business locations with each business categorized into one or more of eight business types: Cafe, Convenience, Gym, Liquor, Lodging, Nightlife, Pharmacy, and Restaurant.   In our supplementary materials, we provide additional details about the processing of each data source and creation of quantitative measures of the surrounding neighborhood context for each vacant lot in the city of Philadelphia.

Before proceeding with our primary matching analysis, we first compare all greened and ungreened vacant lots in Philadelphia in terms of crime rates and surrounding demographic, economic, and land use characteristics.  The substantive differences we observe in surrounding context motivates the need for a careful matching analysis of greened and ungreened vacant lots in Section~\ref{matching}.  

Previous evaluations of this PLC vacant lot greening initiative \citep{branas-1} have employed a difference-in-differences (DID) analysis of crime rates.  In this approach, a difference in crime is calculated for each greened vacant lot as the crime rate over a time period after the greening intervention minus the crime rate over a time period before the greening intervention for that lot.  

In our version of this DID analysis, the time periods are 6 to 18 months before the lot was greened versus 6 to 18 months after the lot was greened.  If we can calculate a corresponding after-before difference in crime outcome for each ungreened vacant lot, then the difference of these greened vs. ungreened differences is the DID estimate of the effect of vacant lot greening on crime rates.   Since the before and after time periods in our analyses are each a one-year span, our DID estimate can be interpreted as the difference in per-year crime rates (i.e. number of crimes per year).  

However, an immediate issue with this approach is that although each greened (``treatment") vacant lot has a well-defined greening intervention date upon which we can center the before vs. after time periods, there is no corresponding intervention date for each of the ungreened (``control") lots in Philadelphia.  We will address this time period issue with our matching analysis in Section~\ref{matching}, but for this preliminary comparison we chose to center our before vs. after time periods for the ungreened vacant lots on October 30th, 2012, which is the average intervention date of the greened vacant lots. 

Figure S1 in our supplementary materials compares the distribution of crime counts around greened lots vs. ungreened lots.  The average after-before reduction in total crimes is -16.0 for greened lots and -44.0 for ungreened lots, which results in a DID estimate of an increase of 28.0 total crimes from vacant lot greening.  However, this simple DID estimate of crime rates does not address the possibility that greened and ungreened vacant lots likely differ in terms of their surrounding context which confounds our attempts to attribute crime differences to the greening intervention.   

Indeed, we observe substantial imbalance between greened and ungreened vacant lots in terms of their surrounding demographic, economic, land use and business characteristics.  In Figure S2 of our supplementary materials, we see that greened and ungreened lots differ substantially in terms of their surrounding racial proportions, per capita income, and proportion of households in each poverty bracket.  In Figures S3-S5 of our supplementary materials, we also see differences between greened and ungreened vacant lots in their surrounding land use zoning, as well as the number and types of businesses.  

These substantial differences in surrounding context for greened versus ungreened vacant lots make it difficult to attribute any observed differences in crime rates to the greening intervention itself.    This problematic imbalance on surrounding characteristics is in addition to the issue that our collection of ungreened vacant (control) lots lack a well-defined intervention date for establishing a before-after change in crime rates.  We address both of these problems with our matching analysis in the next section.

\section{Matched Pairs Comparison of Greened and Ungreened Vacant Lots} \label{matching}

We can address the imbalance in surrounding neighborhood context between greened and ungreened vacant lots by performing a careful matching of each greened (treatment) vacant lot with an ungreened (control) vacant lot that has similar surrounding characteristics.  By creating matched pairs of individual greened and ungreened lots, we can better attribute any observed within-pair difference in crime rates to the greening intervention.  

In addition, our matched pair analysis addresses the issue that there are no intervention dates for our ungreened (control) vacant lots.  Once we have paired up an individual greened vacant lot with a highly similar ungreened vacant lot, we can use the intervention date of that greened lot for its paired ungreened vacant lot.  This ensures that we are comparing changes in crime over the same time period within each of our matched pairs. 

\subsection{Propensity Score Matching} 

In order to create matched pairs of greened vs. ungreened vacant lots with similar surrounding context, we must first choose some measure of the similarity between the surrounding characteristics of any pair of greened and ungreened vacant lots.   We base our matching upon the {\it propensity score} \citep{RosRub83} which is defined as the probability that a particular unit (vacant lot) receives the treatment (greening) based on their surrounding neighborhood context.   

We use a logistic regression model to calculate these propensity scores for each greened and ungreened vacant lot in our data.  In this logistic regression model, each unit $i$ is a vacant lot in the city of Philadelphia with outcome $Y_i = 1$ if vacant lot $i$ was greened or $Y_i = 0$ if vacant lot $i$ was ungreened.  The probability $P(Y_i = 1)$ for each vacant lot $i$ is modeled as a function of its surrounding characteristics $\X_i$ which includes our demographic, economic, land use, and business presence measures outlined in Section~\ref{data-comparisons} and im our supplementary materials.  Details of our fitted logistic regression model are also provided in our supplementary materials.

For each vacant lot $i$ in our data, our fitted logistic regression model produces $\hat{p}_i$ which is the predicted probability of greening for that lot, otherwise known as the {\it propensity score}.  Two vacant lots with highly similar surrounding characteristics will have highly similar propensity scores. We use these propensity scores to match each greened vacant lot to the ungreened vacant lot that has the closest propensity score.   

\begin{figure}[h]
    \includegraphics[width=14cm]{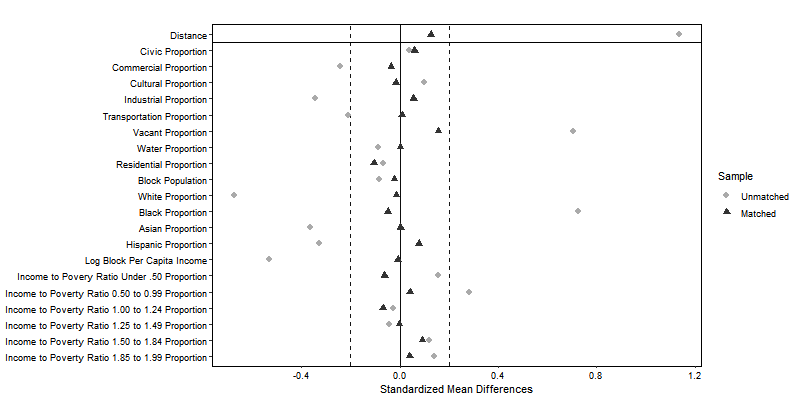}
    \centering
    \caption{Comparing the standardized mean differences in each measure of the surrounding area between greened and ungreened vacant lots before matching (gray dots) and within matched pairs (black triangles).  We observe that the differences between greened and ungreened vacant lots are much closer to zero after matching for nearly all measures of the surrounding area. } \label{fig:a}
\end{figure}

In Figure~\ref{fig:a}, we evaluate the effectiveness of our propensity score matching procedure in terms of improving the similarity (balance) in surrounding context between greened and ungreened lots.  We compare the standardized mean difference in each surrounding area measure between greened and ungreened vacant lots before (gray dots) and after creating matching pairs (black triangles).  

We see that our matched pairs of greened and ungreened lots have much smaller differences on almost all surrounding area measures compared to the sets of all greened and ungreened lots before matching.  
By reducing the average differences in the surrounding characteristics between greened and ungreened vacant lots, our matching procedures allows us to better isolate the effect of the greening intervention on changes in crime rates and how it varies between different nearby land uses and businesses.  

\subsection{Matched Pairs Evaluation of Effect of Greening on Crime} \label{overall-results}

Our evaluation of the effect of the PHS Landcare greening intervention in Section~\ref{data-comparisons} was based on a difference-of-differences (DID) estimate where after-before changes in crime rates were calculated around each vacant lot and then the average differences were calculated between the greened and the ungreened vacant lots.  We correct both the imbalance in surrounding context and the intervention timing issues by now calculating the DID estimate {\it within each matched pair} using the identical time frame (based on the greened lot intervention) for the ungreened and greened lots within each pair.  Our overall estimate of the effect of vacant lot greening on crime is the average of these within-pair DID values.  Since we use one year time periods before and after, the DID estimate can be interpreted as the change in crime rates per year.  

In Table~\ref{tab:h}, we present within-pair DID estimates of the effect of vacant lot greening on crime rates.  We present estimates for total crime as well as the serious and other crime subcategories.  

\begin{table}[h]
\begin{center}
\caption{Within-pair DID estimates of the effect of vacant lot greening on crime rates in a 200 meter radius around each vacant lot. The average pre-greening crime rate for each crime type is provided along with those DID estimates as a percentage of the pre-greening crime rate}\label{tab:h}
\begin{tabular}{lc|ccccc}
\hline
\textbf{Crime} & \textbf{Pre-Greening} & \textbf{DID} & \textbf{Standard}  &  &   & \textbf{DID Estimate} \\ 
\textbf{Category}  & \textbf{Crime Rate} & \textbf{Estimate} & \textbf{Error}  & \textbf{T-stat} & \textbf{p-value}  & \textbf{as a \%}\\ 
\hline
Serious     & 127.8  & -2.71     & 0.46   & -5.87   & 4.76E-09   & -2.1 \% \\ 
Other  & 153.3    & -1.64     & 0.53    & -3.08   & 2.09E-03  & -1.1 \%  \\
Total       & 281.1  & -4.35     & 0.81    & -5.35   & 9.22E-08   & -1.5 \%  \\
\hline
\end{tabular}
\end{center}
\end{table}

The negative signs on the DID estimates imply that the decrease in crime rates (after - before) was larger around the greened vacant lots than their matched ungreened vacant lots. 
We see a beneficial and statistically significant effect of vacant lot greening on the rates of both serious and other crimes.  

In terms of total crime incidence, greened vacant lots showed an additional decrease of 4.35 crimes per year compared to ungreened vacant lots over the same time period, which corresponds to a 1.5\% decrease relative to the pre-greening crime rate.  While a 1-2\% decrease in crime is not a huge effect, it should be noted that this decrease is only for a one year time period and so the long term effect of vacant lot greening could be even more substantial \citep{macdonald2021reducing}.  

Crime rate calculations in Table~\ref{tab:h} were based on a 200 meter radius around each vacant lot (just as in Section~\ref{data-comparisons}) though we also examine results based on 100 and 500 meter radii in our supplementary materials.  We find negative (i.e. a decrease in crime) DID estimates of the effect of vacant lot greening regardless of the choice of radii, and these effects are statistically significant in all cases.  

In the next section, we use our matched pairs to investigate further how the effect of vacant lot greening on crime is potentially moderated by different types of surrounding land use and presence of different business types.

\section{Impact of Land Use Zoning and Business Presence on the Effects of Vacant Lot Greening} \label{land-use-business-results}

Our analysis in Section~\ref{matching} was based on creating matched pairs of individual greened and ungreened vacant lot locations that share highly similar surrounding demographic, economic, land use, and business characteristics.  We can also use these matched pairs to explore whether different types of surrounding land use or business presence are associated with larger or smaller crime effects.  We investigate these associations by subsetting our set of matched pairs into high vs. low levels of certain types of land use zoning or presence vs. absence of certain business types and then comparing the DID estimates of vacant lot greening on crime between these subsets of matched pairs.  

\subsection{Influence of Surrounding Land Use Zoning}

We first investigate whether different types of land use zoning surrounding vacant lots moderates the impact of greening vacant lots on crime.  Specifically, for a particular type of land use zoning designation such as ``commercial", we identify the subsets of our matched pairs with the largest (top 75\%) and the smallest (bottom 25\%) proportions of commercial zoning.  We then calculate DID estimates of the effect of vacant lot greening on crime separately within these two subsets of matched pairs.   Our tabulations of crime are based on a 200 meter radius around each vacant lot, just as in Section~\ref{matching}

In Figure~\ref{fig:impact-plots-combined} (left), we compare our estimated effect of vacant lot greening on total crime between the subsets of matched pairs with the largest (black points) and the smallest proportions (gray points) of each different land use zoning designation described in Section~\ref{data-comparisons}.  For each estimated effect, we also indicate which of these estimates are significantly different from zero (solid points).  For reference, the black vertical line represents the effect of vacant lot greening on total crime across all matched pairs from Table~\ref{tab:h}. 

We see that several types of land use zoning seem to significantly moderate the effect of vacant lot greening on total crime.  Greening of vacant lots in {\it high residential areas} showed an increased reduction on crime compared to areas with low residential proportions.  The greening of vacant lots in {\it high commercial areas} also have a reduction in crime but not as large as seen in high residential areas.  We also see greater reductions in total crime in areas with high proportions of {\it civic/institutions} and low proportions of {\it transportation}. 

\subsection{Influence of Surrounding Business Presence}

We now investigate whether different types of surrounding businesses have an impact on the effect of vacant lot greening on crime.  Specifically, for a particular type of business, such as cafes or restaurants, we identify the subset of our matched pairs where that type of business is present within a 200 meter radius versus the subset of matched pairs where that business is absent within a 200 meter radius.  We then calculate DID estimates of the effect of vacant lot greening on crime separately within these two subsets of matched pairs.  

\begin{figure}[h]
    \includegraphics[width=7cm]{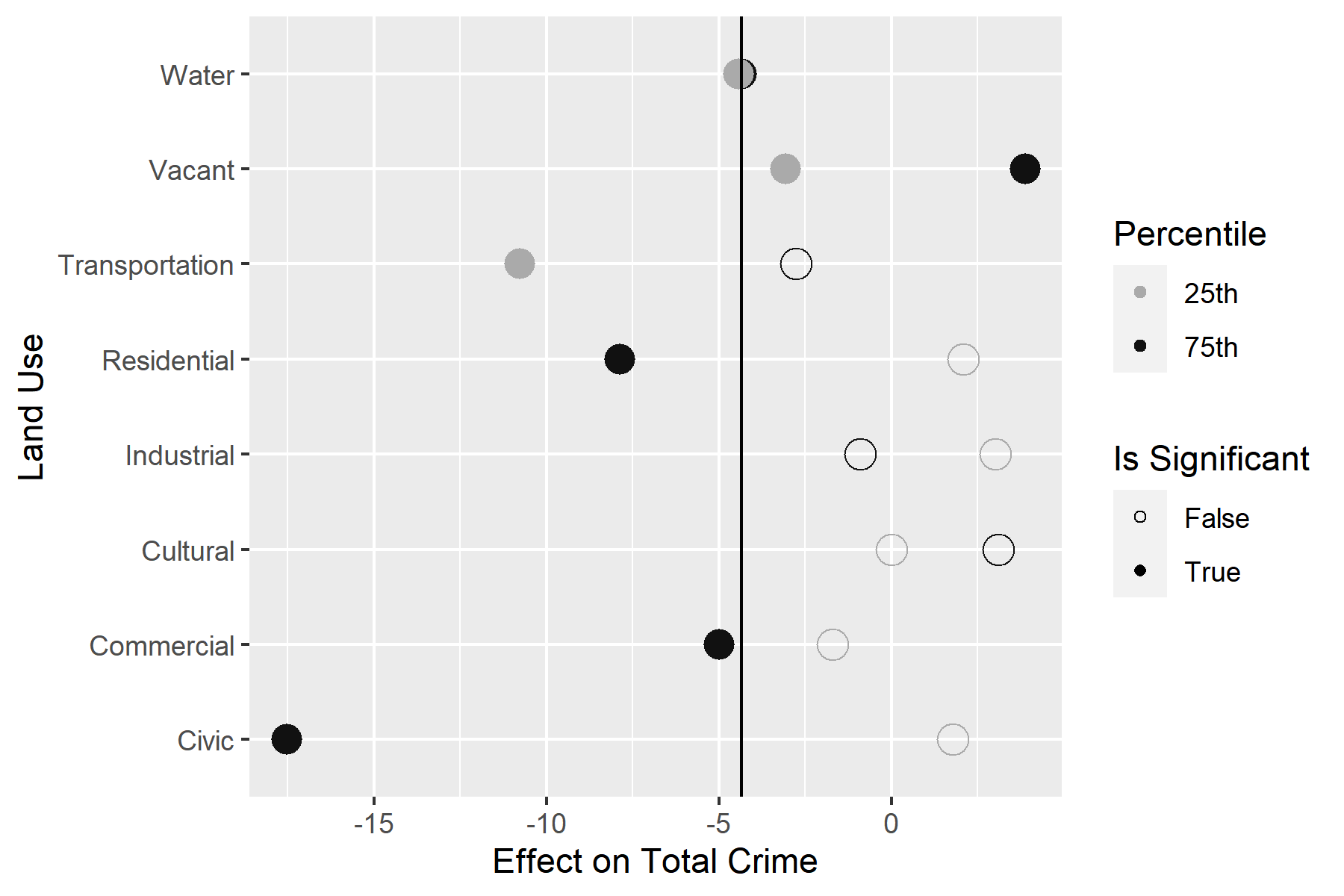} \hfill
    \includegraphics[width=7cm]{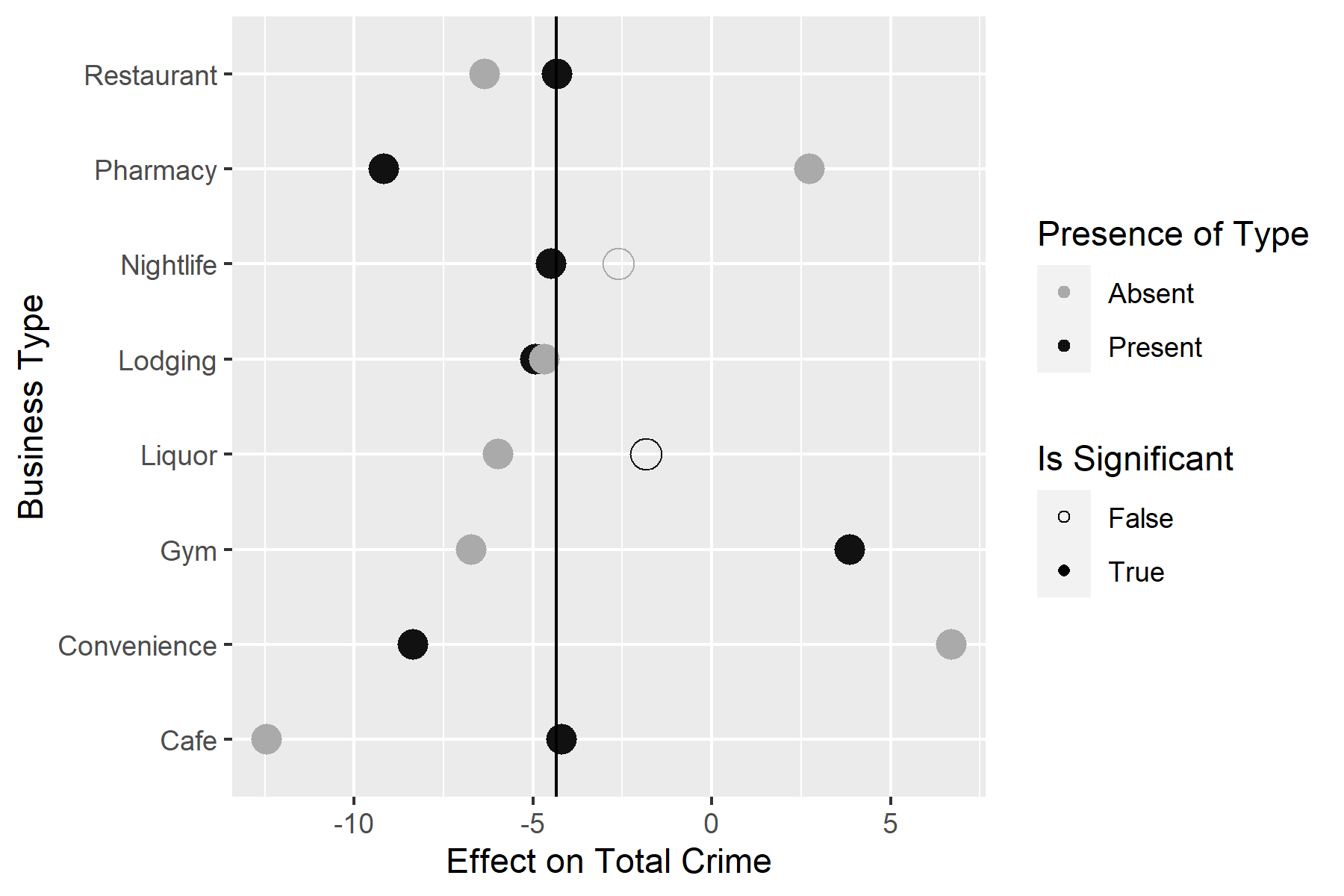}
    \centering
    \caption{{\bf Left:} Comparing vacant lot greening between subsets of matched pairs that differ in terms of a specific land use category.  Each row is a particular land use zoning designation for which we compare two subsets of matched pairs: the black point is the estimated effect of vacant lot greening for matched pairs with the largest (top 25\%) proportion of that land use zoning designation whereas the gray point is the estimated effect of vacant lot greening for matched pairs with the smallest (bottom 25\%) proportion of that designation. {\bf Right:} Comparing vacant lot greening between subsets of matched pairs that differ in terms of presence or absence of a specific business type.  Each row is a particular business type for which we compare two subsets of matched pairs: the black point is the estimated effect of vacant lot greening for matched pairs where that business type is present whereas the gray point is the estimated effect of vacant lot greening for matched pairs where that business type is absent.   {\bf Both plots:} Points are solid circles if the estimated effect is significantly different from zero vs. open circles if not. The black vertical line is the estimated effect of vacant lot greening on total crime across all matched pairs.}    
    \label{fig:impact-plots-combined}
\end{figure}

In Figure~\ref{fig:impact-plots-combined} (right), we compare our estimated effect of vacant lot greening on total crime between the subsets of matched pairs the presence (black points) versus absence (gray points) of each different business type described in Section~\ref{data-comparisons}.   For each estimated effect, we also indicate which of these estimated effects are significantly different from zero (solid points).  The black vertical line again represents the effect of vacant lot greening on total crime across all matched pairs from Table~\ref{tab:h}. 

The presence of convenience stores and pharmacies as well as the absence of cafes, gyms and restaurants are associated with an greater reduction in crime around greened vacant lots.   As we discuss in Section~\ref{discussion}, additional research and new data sources are needed to further investigate these potential associations.

\section{Discussion}\label{discussion}

In this paper, we examine the effect in recent years of the ongoing Philadelphia LandCare (PLC) vacant lot greening program on crime, and assess how surrounding land use and business presence moderates this intervention.  We rely on a propensity score matching analysis that estimates the effect of vacant lot greening on different types of crime while accounting for systematic differences between greened and ungreened lots in terms of their surrounding land use and demographic contexts.   By creating matched pairs of individual greened vs. ungreened vacant lots that share highly similar surrounding characteristics, we can better isolate the effect of the greening intervention on crime and whether it is moderated by nearby land uses and business types.  Our matched pair design also addresses the issue that our ungreened vacant lots (control group) do not have a natural intervention date around which to examine changes in crime.  

Within our matched pairs, we estimate beneficial effects of vacant lot greening on serious crimes, other crimes, and total crimes that correspond to a decrease in the range of 1-2\% reductions per year around greened vacant lots.  These estimated effects of vacant lot greening based on our matched pair analysis suggest more significant decreases in crime than comparisons without matching.  

Our comparisons of subsets of matched pairs that differ substantially on their land use or business characteristics shows the crime reducing effects of vacant lot greening are larger in high residential and low commercial zoning areas, which is consistent with several theories in criminology and urban planning \citep{CohFel79,Tay88,WilCul18}.  The increased reduction in crime in more residential areas may indicate higher guardianship or collective efficacy among residents when lots are greened nearby their homes \citep{Tay88, Sam12}. 

We also see that the greening of vacant lots was associated with greater reductions in total crime in areas with high proportions of {\it civic/institutions} and low proportions of {\it transportation}. These findings are consistent with theory that suggests places with more human activity and anonymity are more prone to crime, and places with greater place management are less likely to have crime \citep{WilCul18}.

We also find that the crime reducing effects of vacant lot greening are larger when pharmacies and convenience stores are present nearby as well as in the absence of gyms, cafes, and restaurants. These findings provide empirical support for crime opportunity theories \citep{WilCul18} that would argue that busier streets that encourage crime can be offset when business operators act as guardians or place managers, which is an idea that also dates back to Jane Jacobs \citep{Jac61}. It is plausible that pharmacies and convenience stores are more concerned with keeping greened vacant lots nearby their businesses free of crime because it may discourage people from shopping at those locations. 

However, further research is needed in order to confirm these potential associations and investigate the underlying mechanisms by which the built environment impacts safety.  In particular, it is important to assess the dynamic changes in when businesses open and close and to measure human activity around vacant lots.  
There is also a need for more research on the impact of vacant lot greening initiatives on outcomes beyond crime and safety.  \cite{heckert_mennis-1} found that property values surrounding greened vacant lots had a greater increase in value than properties surrounding non-greened vacant lots.   \cite{branas-1} found that vacant lot greening was associated with residents reporting less stress and more exercise in certain areas of Philadelphia.  \cite{south-1} found that views of a greened vacant lot was associated with a significant reduction in heart rate and concluded that remediating neighborhood blight may reduce stress and improve health. Feelings of depression and self-reported poor mental health were reduced in participants living near greened vacant lots \citep{south-2}. 

Nevertheless, our matching analyses indicate promising results that highlight how the PLC greening intervention continues to be associated with significant reductions in crime in more recent years, and that the effects are larger when lots are greened in residential areas, around civic institutions, and nearby to pharmacies and convenience stores.  These findings suggest that greening interventions may want to be prioritized when they are located nearby residents or businesses that can act as proprietors of the space and help amplify the crime reducing benefits of remediating vacant lots. This research can be used to better inform the replications of the PLC program in other cities as a potential benefit to improving safety in urban areas.  

\section{Acknowledgements} 

We thank Ajjit Narayanan and Park Sincharisi for helpful contributions to this research.  We are also grateful to the Wharton Social Impact Initiative for their generous support of our work. We owe special thanks to the Pennsylvania Horticultural Society for their collaboration and providing access to the data on location and timing of their vacant lot greening interventions.  

\bibliographystyle{SageH}
\bibliography{references}

\newpage

\setcounter{section}{0}

\begin{center}
{\LARGE {\bf Supplementary Materials for ``The Effects of Vacant Lot Greening and the Impact of Land Use and Business Presence on Crime"}}
\end{center}

\bigskip

\section{Urban Data in Philadelphia} \label{data}

Our analyses will be based on publicly available data on crime, economic and demographic neighborhood characteristics, and land use zoning as well as a comprehensive database on business locations and open hours for the city of Philadelphia that has been compiled by our research group.  We have the location and type of every reported crime over the past decade from the Philadelphia Police Department. We also have detailed data on neighborhood-level income, poverty, race and population density from the U.S. Census Bureau. The city of Philadelphia provides zoning designation for the approximately half million lots in the city, which we use to summarize the land use around vacant lots.  Below we provide additional details about the processing of each data source and creation of quantitative measures of the surrounding area for each vacant lot in the city of Philadelphia.

\subsection{Vacant Lots Data}

We have data on the location (and date of greening) for each vacant lot greened through the Pennsylvania Horticultural Society's Land Care program.  We focus on 4651 vacant lots greened in the period from 9/01/2007 - 9/01/2017, which is the time period for which we also have high resolution crime data (Section~\ref{crimedata}).  We also have data on the location of ungreened vacant lots over the same time period from the City of Philadelphia's Licenses and Inspections Office.   We filtered this data to only include vacant property (non-building) violations and removed duplicate violations at the same location by only including the first violation instance at each vacant lot location.   After this filtering, we have the locations of $\approx$ 16800 ungreened vacant lots in Philadelphia.  

\subsection{Crime Data} \label{crimedata}

We retrieved crime data for the city of Philadelphia that is made available by the Philadelphia Police Department on the opendataphilly.org data portal.  Our dataset contains the date, time and GPS location of each reported crime from 2007 to 2019, as well as the type of crime (e.g. homicide, aggravated assault, etc).   We filtered out some infrequent crime types that are unlikely to be related to the use of public spaces (such as fraud, embezzlement and offenses against the family and children).  After this filtering, we have $\approx 1.5$ million reported crimes in the 2007-2019 time period.  We categorized these crimes into two major types: \textbf{serious crimes} versus \textbf{other crimes}.  \textbf{Serious crimes} include homicides, rapes, robberies, and both aggravated and other assaults.  \textbf{Other crimes} are a composite of property and nuisance crimes, which includes the specific categories of burglaries, thefts, vehicle thefts, vandalism, public drunkenness, disorderly conduct, and vagrancy/loitering.  

\subsection{Demographic Data} \label{demodata}

Population demographic data for Philadelphia was obtained from the U.S. Census Bureau website (Table SF1 P5 in their data portal).  The raw demographic data gives the population count by race from the 2010 census in each of the 18,872 census blocks in Philadelphia.  We use this data to calculate the population count and racial proportions (Black, White, Hispanic, and Asian) surrounding each greened and ungreened vacant lot.  

\subsection{Economic Data}

Economic data for Philadelphia was obtained from the 2015 American Community Survey (Tables B19301 and C17002 in the U.S. Census Bureau data portal).  This data contains the per-capita mean income and the proportion of households in seven different poverty brackets based on the ratio of income to poverty line for each census block group in Philadelphia.  Specifically, the seven poverty brackets are: [0, 0.5), [0.5, 1), [1, 1.25), [1.25, 1.5), [1.5, 1.85), [1.85, 2) and [2, $\infty$) where, for example, the [0.5, 1) bracket is households with income between 50\% of the poverty line and the poverty line.

\subsection{Land Use Zoning Data} \label{landusedata}

Zoning data is made available by the City of Philadelphia through the opendataphilly.org data portal.  This data consists of a shapefile that provides the area and registered land use zoning designation for the $\approx 560,000$ lots in the city.  We aggregated these zoning designations into eight primary types: Residential, Commercial, Industrial, Civic (also includes Institutions), Transportation, Cultural (also includes Parks), Water, and Vacant

We use these registered zoning designations to create several quantitative measures of the land use around each of the greened and ungreened vacant lot locations in Philadelphia.   Specifically, for the area in a 200 meter radius around each vacant lot location, we calculate the proportion of that area that is designated as each of those eight zoning types, i.e. the proportion of residential land use, proportion of commercial land use, etc. 

\subsection{Business Data} \label{busdata}

Our research group manually assembled a database of Philadelphia business locations by scraping three different web resources (Google Places, Yelp, and Foursquare).   Each business is categorized into one or more of eight business types: Cafe, Convenience, Gym, Liquor, Lodging, Nightlife, Pharmacy, and Restaurant.  This data is described in more detail in \cite{humphrey2020urban}.  

From this database, we create measures of business presence around each of the greened and ungreened vacant lot locations in Philadelphia.
For each vacant lot, we create a set of eight binary variables (for our eight business types) that indicate whether there is a business of that particular type present within 200 meters of that vacant lot.   For each vacant lot, we also tabulate the total number of businesses located within 200 meters of that vacant lot.

\section{Comparisons between Unmatched Greened and Ungreened Vacant Lots}

\begin{figure}[ht]
\renewcommand\thefigure{S1}
\includegraphics[width=12cm]{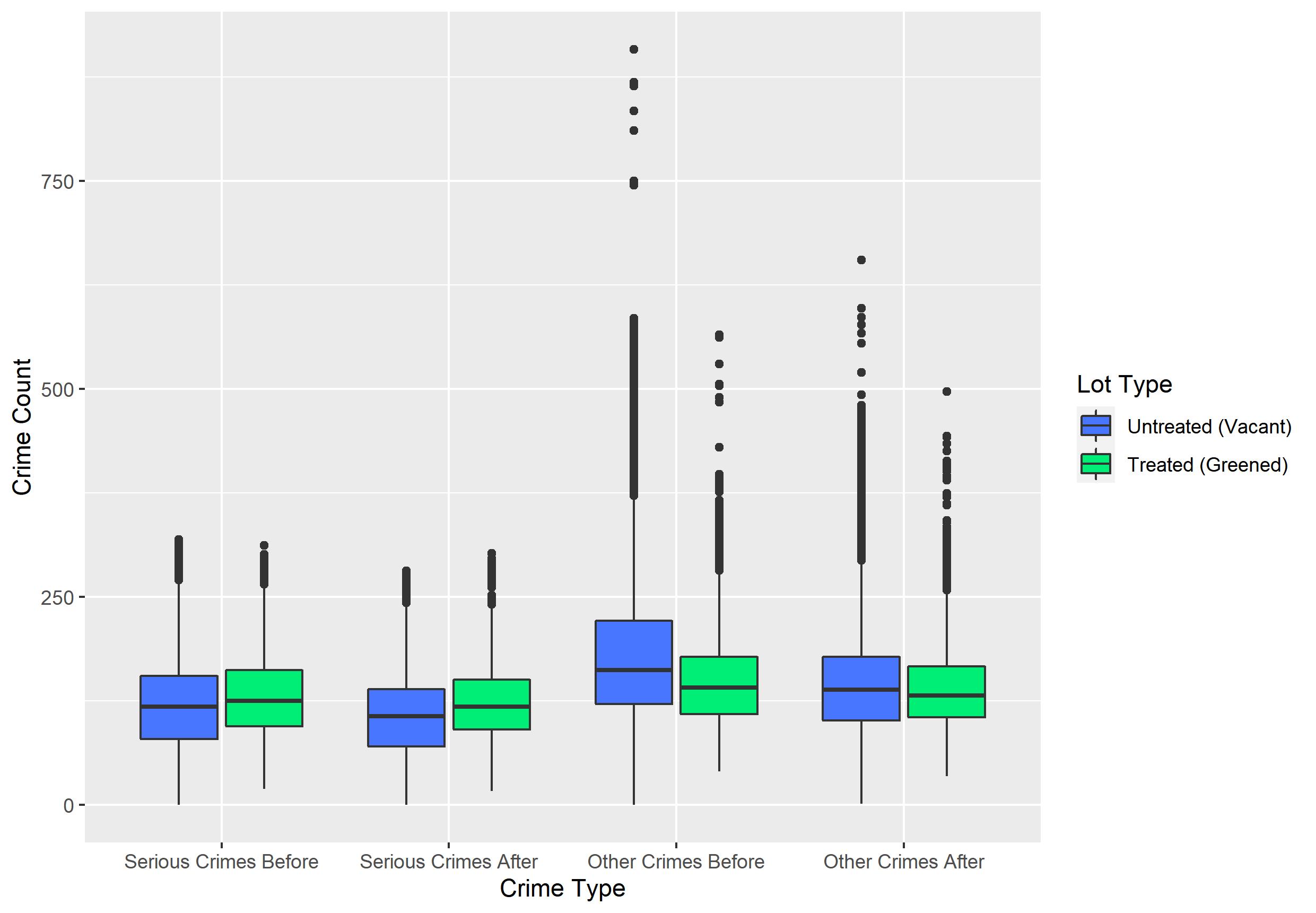}
\centering
\caption{Distribution of serious crime counts and other crime counts within 200m radius of all greened and ungreened vacant lot locations in Philadelphia} \label{fig:crimecomparison-unmatched}
\end{figure}

Figure~\ref{fig:crimecomparison-unmatched} compares the distribution of crime counts in a 200 meter radius around greened lots vs. ungreened lots in the time periods before and after either the greening intervention for that lot (in the case of greened lots) or October 30th, 2012 (in the case of ungreened lots).  We observe similar patterns for greened and ungreened vacant lots: a decrease in crime counts in the ``after" time period compared to the ``before" crime period.   The average after-before reduction in total crimes is -16.0 for greened lots and -44.0 for ungreened lots, which results in a DID estimate of an increase of 28.0 total crimes from vacant lot greening.  This DID analysis clearly does not provide evidence for beneficial effects of vacant lot greening on crime.  

\begin{figure}[ht]
\renewcommand\thefigure{S2}
\includegraphics[width=12cm]{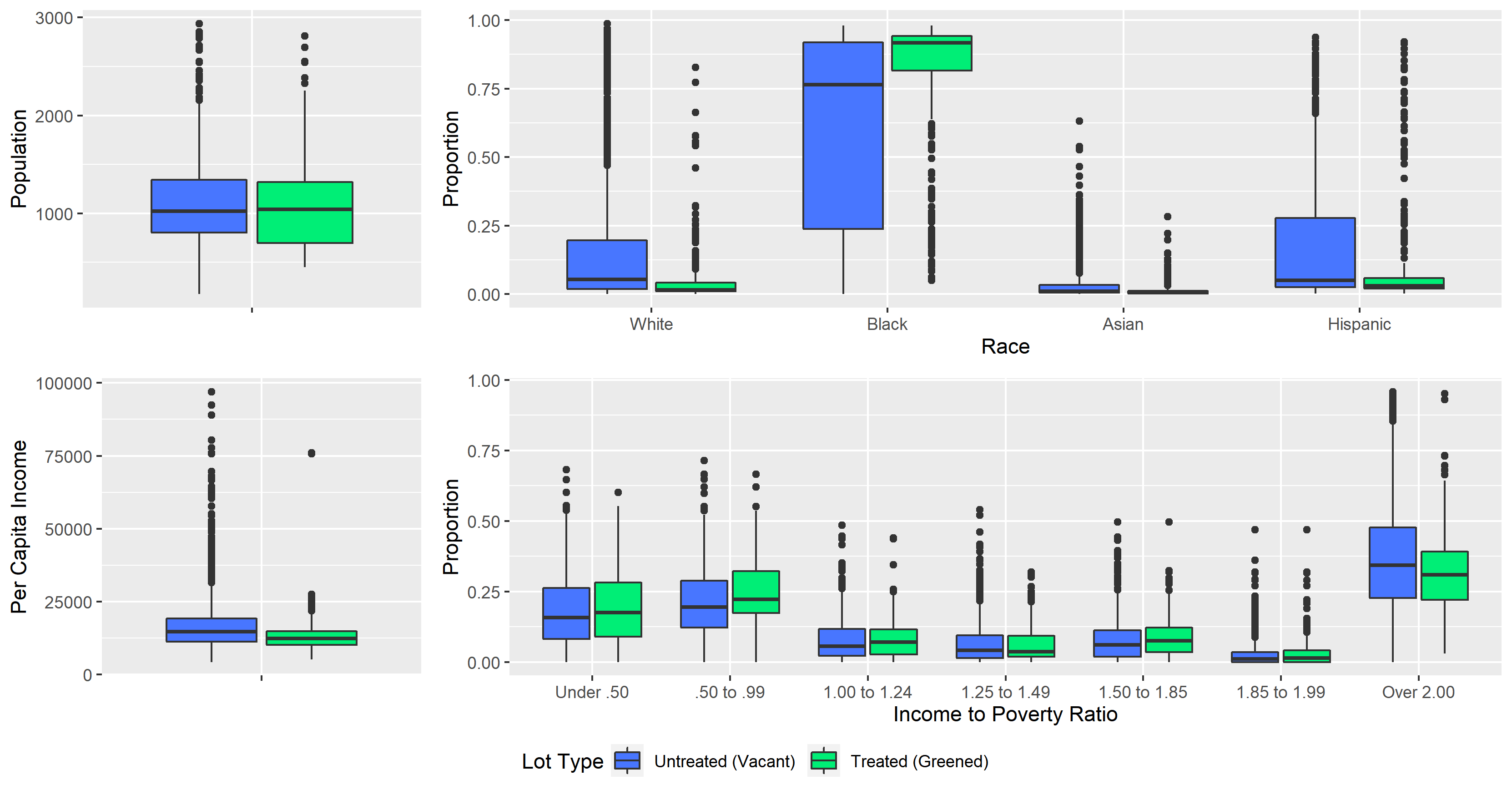}
\centering
\caption{Boxplots that compare greened versus ungreened vacant lots in terms of population count (top left), racial proportions (top right), per capita income (bottom left) and poverty brackets (bottom right). The central black line in each boxplot is the median value and the width of the box is the inter-quartile range.}.
\label{fig:covariate-imbalance}
\end{figure}

Figure~\ref{fig:covariate-imbalance} gives side-by-side boxplots that compare greened vacant lots to ungreened vacant lots on several demographic and economic measures.  

We see that greened and ungreened lots differ substantially in terms of their surrounding racial proportions, per capita income, and proportion of households in each poverty bracket.   We observe substantially lower proportions of Hispanics and Whites and substantially higher proportions of Blacks in the neighborhoods surrounding greened lots compared to ungreened lots.   The neighborhoods surrounding greened lots also tend to have lower per capita income and a larger proportion of households in the high poverty brackets.  We also see differences between greened and ungreened vacant lots in terms of the surrounding land use zoning, as well as number and types of businesses.  Details on this comparison are given in our supplementary materials.  

\begin{figure}[ht!]
\renewcommand\thefigure{S3}
    \includegraphics[width=12cm]{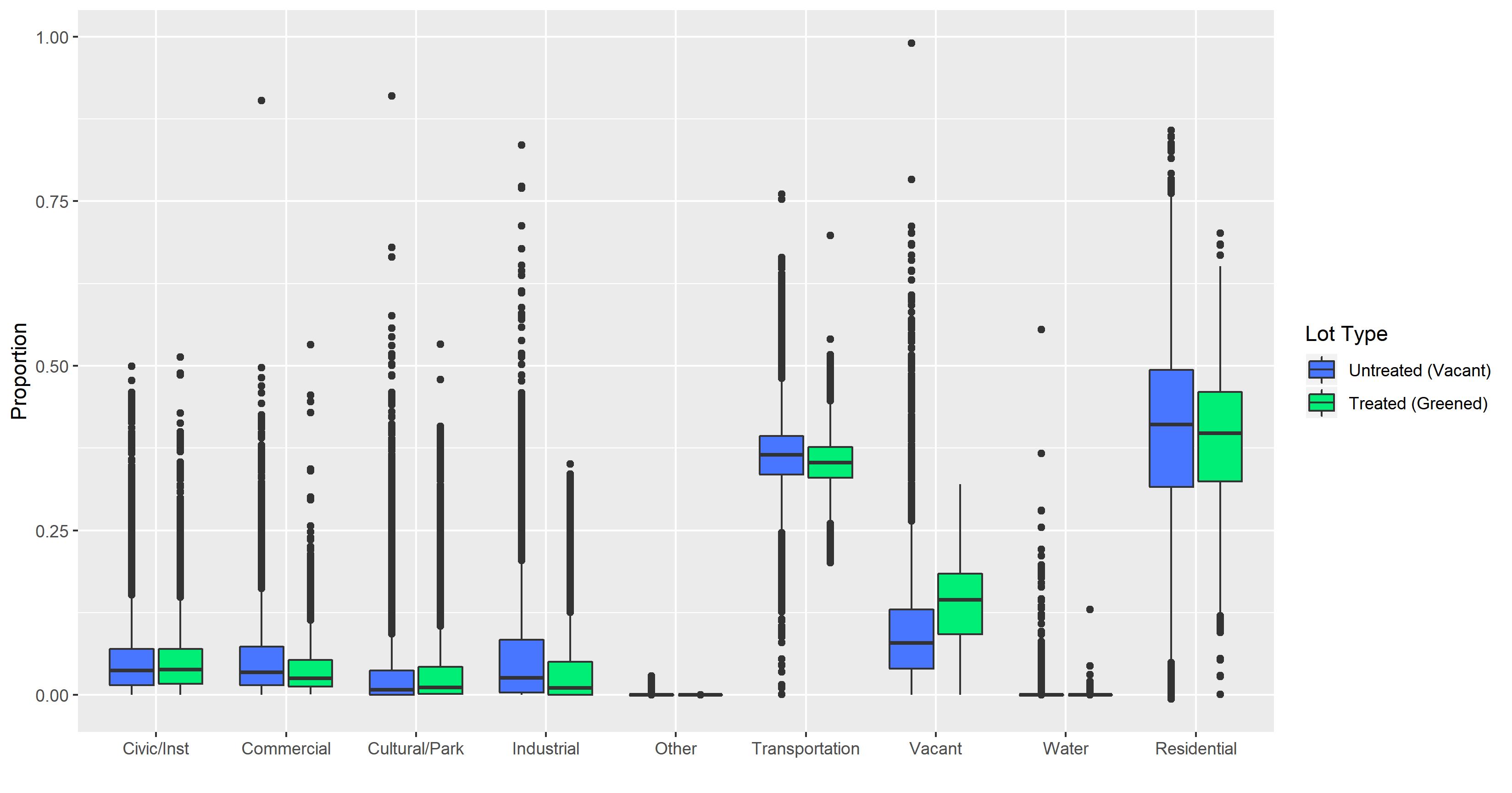}
    \centering
    \caption{Land use proportions (based on zoning data) for greened versus ungreened vacant lots}
    \label{fig:figure7}    
    \end{figure}

Figure \ref{fig:figure7} displays side-by-side boxplots comparing greened vs. ungreened vacant lots in terms of land use proportions that we created from the City of Philadelphia zoning data. 

We observe that greened vacant lots tend to occur in neighborhoods with less commercial and industrial zones but more cultural/park zones compared to ungreened vacant lots.   The variance of transportation land use around greened lots is smaller than the variance of transportation land use around ungreened lots. We also observe that greened lots tend to have a higher proportion of surrounding vacant land use than ungreened vacant lots.
    
\begin{figure}[ht!]
\renewcommand\thefigure{S4}
    \includegraphics[width=10cm]{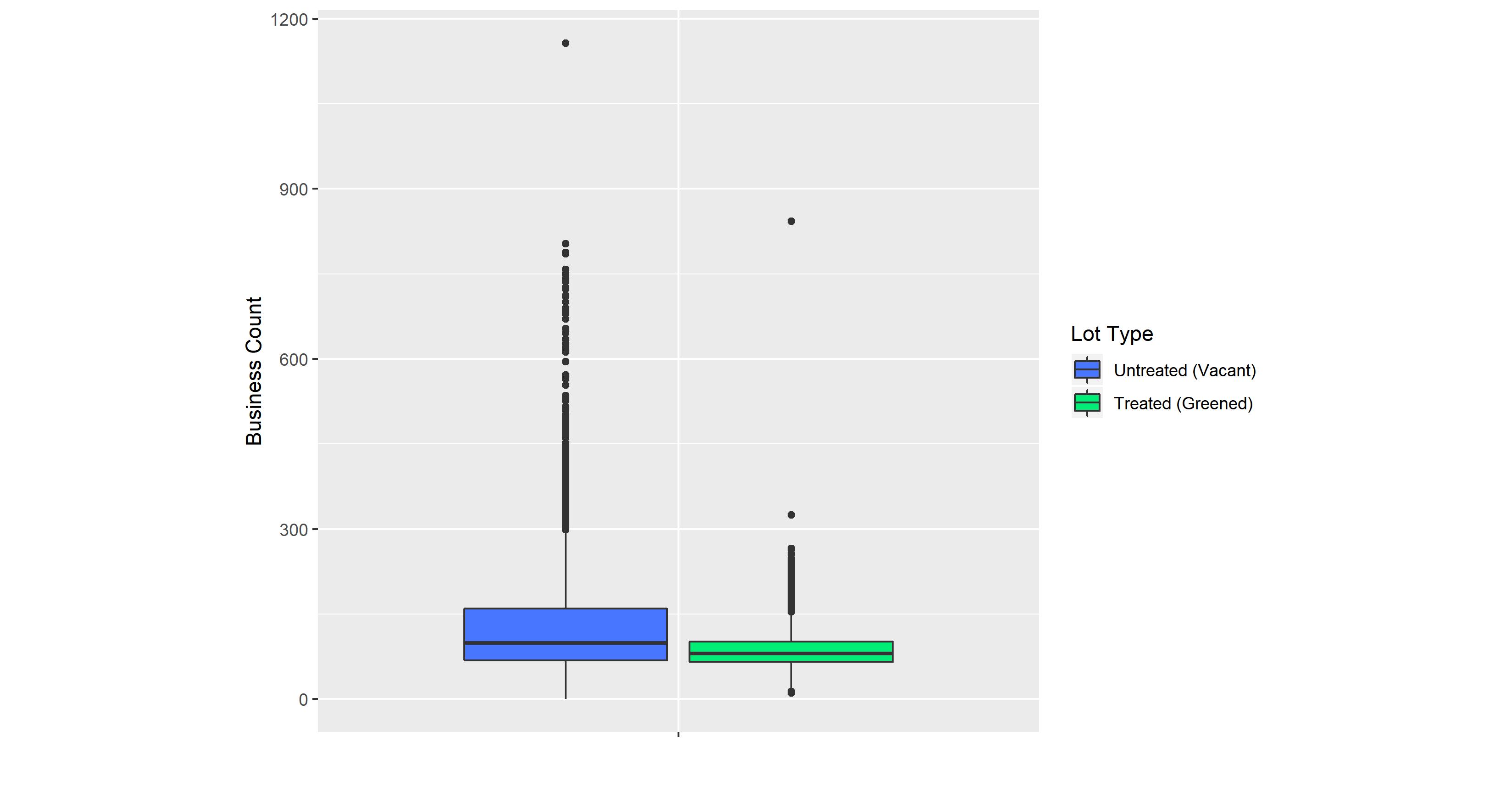}
    \centering
    \caption{Number of Businesses surrounding greened vs. ungreened vacant lots}
    \label{fig:figure8}
    \end{figure}

Figure \ref{fig:figure8} displays side-by-side boxplots comparing greened vs. ungreened vacant lots in terms of the number of businesses in a 200 meter radius. We observe a smaller number of businesses surrounding greened lots compared to the number of businesses surrounding ungreened vacant lots, which is also evident in the lower commercial proportion around greened lots seen in Figure \ref{fig:figure7}.

\begin{figure}[ht!]
\renewcommand\thefigure{S5}
    \includegraphics[width=10cm]{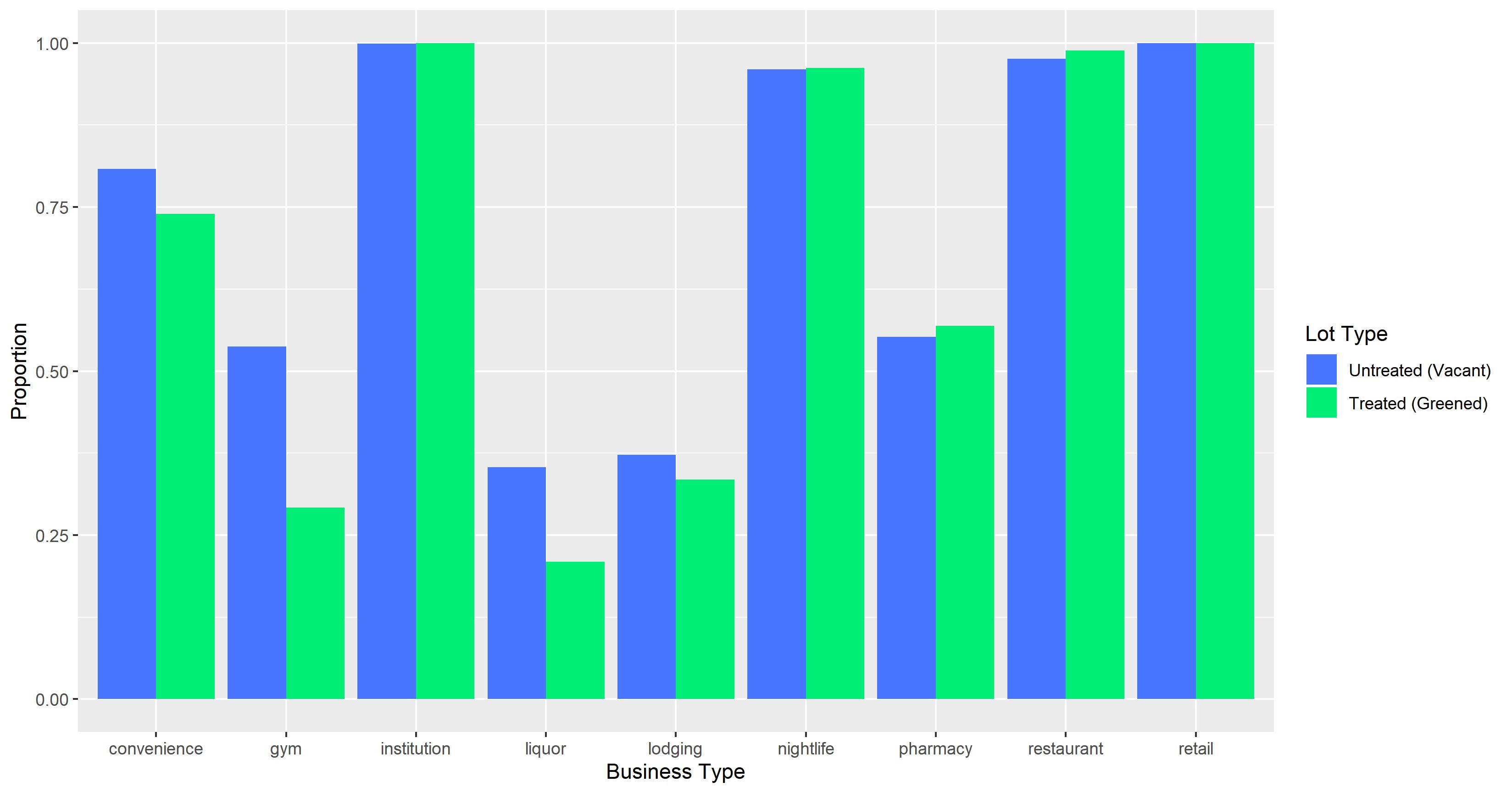}
    \centering
    \caption{Proportions of each business type surrounding of greened and non-greened vacant lots}
    \label{fig:figure9}
    \end{figure}

Figure \ref{fig:figure9} gives barplots that compare greened and ungreened vacant lots in terms of the proportions of each business type within a 200 meter radius.  We observee that greened vacant lots have a lower proportion of convenience stores, gyms, liquor stores compared to ungreened vacant lots.

\section{Logistic Regression Model for Propensity Scores}

We used a logistic regression model to calculate the propensity scores for each greened and ungreened vacant lot in our data.  In this logistic regression model, each unit $i$ is a vacant lot in the city of Philadelphia with outcome $Y_i = 1$ if vacant lot $i$ was greened or $Y_i = 0$ if vacant lot $i$ was ungreened.  The probability $P(Y_i = 1)$ for each vacant lot $i$ is modeled as a function of its surrounding characteristics $\X_i$ which includes our demographic, economic, land use and business presence measures outlined in the data section of our paper.

The “Other” land use zoning proportion and one poverty bracket (income to poverty line above 2.00) were removed from the model due to high collinearity with the other surrounding characteristics.  We also removed the indicators for retail business from the model since almost all vacant lots had at least one of this type of business in their surrounding area.  

In Table~\ref{tab:perfm}, we provide several common evaluation metrics of our fitted logistic regression model.  For the ``Accuracy" and ``Balanced Accuracy" metrics, we chose the decision boundary to be the proportion ($p = 0.22$) of all vacant lots in our dataset that are greened.  We compare the fitted logistic regression model that uses all available measures of the surrounding area (``All") to fitted logistic regression models that only use one set of measures (e.g. ``Economic" vs. ``Demographic").  

\begin{table}[ht]
\renewcommand\thetable{S1}
\begin{center}
\caption{Evaluation Metrics of Logistic Regression Models for different sets of included surrounding characteristics. } \label{tab:perfm}
\begin{tabular}{@{}lccccc@{}}
\toprule
\textbf{Metric}               & \textbf{All} & \textbf{Economic} & \textbf{Demographic} & \textbf{Land Use} & \textbf{Business} \\ \midrule
\textbf{ROC AUC}              & 0.81          & 0.66           & 0.71              & 0.74           & 0.67           \\
\textbf{Accuracy}             & 0.71          & 0.61           & 0.59              & 0.68           & 0.61           \\
\textbf{Balanced Accuracy}    & 0.74          & 0.63           & 0.65              & 0.68           & 0.64           \\
\textbf{Kappa}                & 0.37          & 0.18           & 0.20              & 0.28           & 0.20           \\
\textbf{Sensitivity}          & 0.69          & 0.59           & 0.55              & 0.67           & 0.59           \\
\textbf{Specificity}          & 0.79          & 0.66           & 0.74              & 0.69           & 0.70           \\
\textbf{Pos Pred Value}       & 0.92          & 0.86           & 0.89              & 0.89           & 0.88           \\
\textbf{Neg Pred Value}       & 0.42          & 0.31           & 0.31              & 0.37           & 0.32           \\
 \bottomrule
\end{tabular}
\end{center}
\end{table}

We can see that the logistic regression model that uses all surrounding characteristics has the best fit to the data by all evaluation metrics, which suggests that each data type is making a significant contribution to the model.  Amongst the models based just on single data type, the land use zoning characteristics seem to provide the best fit to the data.  

\begin{figure}[ht]
\renewcommand\thefigure{S6}
    \includegraphics[width=6cm]{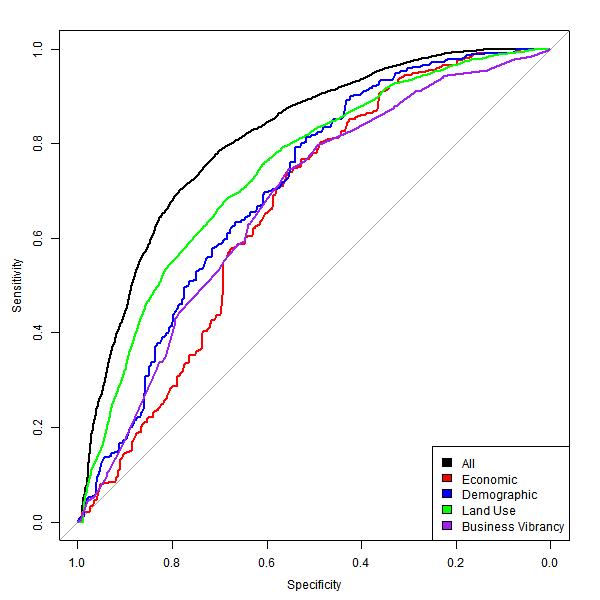}
    \centering
    \caption{ROC Curves for Different Covariate Groups Predicting Greening in Lots}
    \label{fig:figure10}
\end{figure}

These observations are confirmed when we compare the ROC (Receiving Operating Characteristic) curves for these different fitted logistic regression models in Figure \ref{fig:figure10}.  We see that the model using ``All" surrounding characteristics has the best ROC curve, followed by the model that only uses the surrounding land use zoning proportions. 

In Table~\ref{tab:a}, we examine the estimated coefficients from the fitted logistic regression model that uses all surrounding characteristics.   We see that coefficients with the largest (in magnitude) statistics are income per capita, the racial proportions, population count and indicators for several of the business types.  

\begin{table}[ht]
\renewcommand\thetable{S2}
\begin{center}
\caption{Summary of Coefficients from Logistic Regression Model using all surrounding characteristics}\label{tab:a}
\begin{tabular}{@{}lrrrr@{}}
\toprule
\textbf{Coefficient}         & \textbf{Estimate} & \textbf{Std. Error} & \textbf{Stat} & \textbf{P-value} \\ \midrule
\textbf{(Intercept)}         & -2007.95      & 1132.89         & -1.77     & 0.08                    \\
\textbf{Civic}               & 0.63       & 0.36         & 1.73      & 0.08                    \\
\textbf{Commercial}          & 1985.41       & 1116.17           & 1.78      & 0.08                     \\
\textbf{Cultural}            & 1988.93       & 1116.15          & 1.78      & 0.07                    \\
\textbf{Industrial}          & 1983.26       & 1116.16         & 1.78      & 0.08                    \\
\textbf{Transportation}      & 1984.24        & 1116.14         & 1.78      & 0.08                    \\
\textbf{Vacant}              & 1993.15      & 1116.15          & 1.79       & 0.07                    \\
\textbf{Water}               & 1968.91       & 1116.14         & 1.76      & 0.08                     \\
\textbf{Residential}         & 1985.70       & 1116.15         & 1.78      & 0.08                     \\
\textbf{block\_total\_count} & 2.37E-04       & 4.67E-05            & 5.08      & 3.80E-07                       \\
\textbf{white\_percent}      & 9.55       & 2.33         & 4.13      & 3.70E-05                       \\
\textbf{black\_percent}      & 12.41       & 2.25         & 5.52      & 3.42E-08                       \\
\textbf{asian\_percent}      & 12.19       & 2.39         & 5.10      & 3.35E-07                       \\
\textbf{hispanic\_percent}   & 10.26       & 2.21         & 4.64      & 3.56E-06                       \\ 
\textbf{block\_per\_capita\_income}          & -1.47      & 0.10         & -14.21     & 7.61E-46                       \\
\textbf{income\_to\_poverty\_under\_.50}     & -1.65       & 0.27         & -6.10     & 1.07E-09                       \\
\textbf{income\_to\_poverty\_.50\_to\_.99}   & 0.08       & 0.25         & 0.31      & 0.76                    \\
\textbf{income\_to\_poverty\_1.00\_to\_1.24} & -2.24      & 0.34         & -6.67     & 2.50E-11                       \\
\textbf{income\_to\_poverty\_1.25\_to\_1.49} & -0.70       & 0.34         & -2.11      & 0.04                    \\
\textbf{income\_to\_poverty\_1.50\_to\_1.84} & 1.46       & 0.32         & 4.61      & 4.01E-06                       \\
\textbf{income\_to\_poverty\_1.85\_to\_1.99} & 0.58       & 0.36         & 1.61       & 0.11                    \\
\textbf{cafe}                                & 0.28       & 0.09         & 2.97      & 3.03E-03                    \\
\textbf{convenience}                         & -0.06      & 0.05         & -1.20     & 0.23                    \\
\textbf{gym}                                 & -0.51      & 0.04         & -11.71     & 1.08E-31                       \\
\textbf{institution}                         & 12.02       & 131.09          & 0.09      & 0.93                    \\
\textbf{liquor}                              & -0.29      & 0.05         & -6.31     & 2.76E-10                       \\
\textbf{lodging}                             & -0.02      & 0.04         & -0.43     & 0.66                    \\
\textbf{nightlife}                           & 0.41       & 0.10         & 4.02      & 5.90E-05  \\
\textbf{pharmacy}                            & 0.41       & 0.04          & 9.67      & 3.87E-22                       \\
\textbf{restaurant}                          & 1.03       & 0.17         & 6.23      & 5.03E-10                       \\
\textbf{retail}                              & 9.71       & 143.07          & 0.07      & 0.95                    \\
\textbf{business\_count}                     & 2.77E-03      & 5.22E-04         & -5.31     & 1.11E-07     \\\bottomrule
\end{tabular}
\end{center}
\end{table}

\newpage

\section{DoD Estimates using Different Radii around Vacant Lots}

In Table 1 of our main paper, we present within-pair DID estimates of the effect of vacant lot greening on crime where crime rates were calculated based on a 200 meter radius around each vacant lot.  In this section, we examine the within-pair DoD estimates of the effect of vacant lot greening on crime based on alternative radii of 100 meters and 500 meters.  In Table~\ref{tab:100}, we present within-pair DID estimates of the effect of vacant lot greening on crime using a radius of 100 meters around each vacant lot.  In Table~\ref{tab:500}, we present within-pair DID estimates of the effect of vacant lot greening on crime using a radius of 500 meters around each vacant lot.  

Comparing to the 200 meter results in Table 1 of our main paper, we are reassured to see in Table~\ref{tab:100} and Table~\ref{tab:500} that all difference-in-difference estimates are negative (i.e. a reduction in crime) and statistically significant regardless of the choice of radii.   In comparing the DoD estimates as percentage changes, we see reductions in the 1-2\% range when using 500 meters as the radius which is highly similar to our results using 200 meters.  We see even larger percentage changes when using a radius of 100 meters, but we are cautious that these larger percentage effects are, in part, driven by the smaller crime counts that are observed when using this smaller radius.  For these reasons, we focus on the results for 200 meters in our main paper.  

\begin{table}[ht]
\renewcommand\thetable{S3}
\begin{center}
\caption{Within-pair difference-in-difference estimates of the effect of vacant lot greening on crime rates using a 100 meter radius. We also provide the average pre-greening crime rate for each crime type as well as difference-in-differences estimates as a percentage of that pre-greening crime rate}\label{tab:100}
\begin{tabular}{lc|ccccc}
\hline
\textbf{Crime} & \textbf{Pre-Greening} & \textbf{DID} & \textbf{Stand.}  &  &   & \textbf{DID Est.} \\ 
\textbf{Type}  & \textbf{Crime Rate} & \textbf{Estimate} & \textbf{Error}  & \textbf{T-stat} & \textbf{p-value}  & \textbf{as a \%}\\ 
\hline
Serious & 33.9  & -1.19     & 0.24   & -4.93   & 8.4E-07   & -3.5 \% \\ 
Other   & 39.7  & -1.02     & 0.24    & -4.26   & 2.1E-05  & -2.6 \%  \\
Total   & 73.7  & -2.21     & 0.38    & -5.81   & 6.8E-09   & -3.0 \%  \\
\hline
\end{tabular}
\end{center}
\end{table}

\begin{table}[ht]
\renewcommand\thetable{S4}
\begin{center}
\caption{Within-pair difference-in-difference estimates of the effect of vacant lot greening on crime rates using a 500 meter radius. We also provide the average pre-greening crime rate for each crime type as well as difference-in-differences estimates as a percentage of that pre-greening crime rate}\label{tab:500}
\begin{tabular}{lc|ccccc}
\hline
\textbf{Crime} & \textbf{Pre-Greening} & \textbf{DID} & \textbf{Stand.}  &  &   & \textbf{DID Est.} \\ 
\textbf{Type}  & \textbf{Crime Rate} & \textbf{Estimate} & \textbf{Error}  & \textbf{T-stat} & \textbf{p-value}  & \textbf{as a \%}\\ 
\hline
Serious & 674.0  & -7.22     & 1.11   & -6.49   & 9.4E-11   & -1.1 \% \\ 
Other   & 919.4    & -18.59     & 1.54    & -12.08   & 4.7E-33  & -2.0 \%  \\
Total   & 1593.4  & -25.81     & 2.16    & -11.92   & 2.8E-32   & -1.6 \%  \\
\hline
\end{tabular}
\end{center}
\end{table}

\end{document}